\documentclass[conference]{IEEEtran}
\IEEEoverridecommandlockouts
\usepackage{cite}
\usepackage{amsmath,amssymb,amsfonts}
\usepackage{graphicx}
\usepackage{textcomp}
\usepackage{xcolor}
\usepackage{listings}  
\usepackage{xcolor}
\usepackage{algorithm}
\usepackage{algpseudocode, cleveref}
\usepackage{subcaption}

\def\BibTeX{{\rm B\kern-.05em{\sc i\kern-.025em b}\kern-.08em
    T\kern-.1667em\lower.7ex\hbox{E}\kern-.125emX}}
\begin{document}

\title{Quantum Circuit Partitioning For Effective Utilization of Quantum Resources\\
\thanks{Corresponding author: vardaan.s@thewiser.org}
}

\author{\IEEEauthorblockN{Connor Howe}
\IEEEauthorblockA{\textit{The Washington Institute for STEM} \\
\textit{Entrepreneurship and Research (WISER)}\\
Washington DC, USA \\
0009-0004-0743-2479}
\and
\IEEEauthorblockN{Cristina Radian}
\IEEEauthorblockA{\textit{The Washington Institute for STEM} \\
\textit{Entrepreneurship and Research (WISER)}\\
Washington DC, USA \\
0009-0006-3642-5416}
\and
\IEEEauthorblockN{Justin Woodring}
\IEEEauthorblockA{\textit{The Washington Institute for STEM} \\
\textit{Entrepreneurship and Research (WISER)}\\
Washington DC, USA \\
0000-0002-0765-5654}
\and
\IEEEauthorblockN{Vardaan Sahgal}
\IEEEauthorblockA{\textit{The Washington Institute for STEM} \\
\textit{Entrepreneurship and Research (WISER)}\\
Washington DC, USA \\
0000-0002-2293-7430}
\and
\IEEEauthorblockN{ Brian J. McDermott}
\textit{Naval Nuclear Laboratory}\\
Niskayuna, USA \\
0000-0002-0335-9508}

\maketitle

\begin{abstract}

    Near-term hardware is constrained by high error rates, small qubit counts, and relatively low output fidelity, making the execution of large, high performance quantum circuits difficult. Circuit partitioning (or circuit cutting) has emerged as a promising approach to circumvent these limitations by decomposing circuits into smaller subcircuits at two-qubit interaction points. However, it remains unclear which classes of circuits benefit the most from partitioning and under what hardware conditions it is most effective. In this work, we evaluate the suitability of quantum circuits for partitioning from three perspectives: improving fidelity, enabling distributed execution, and scaling to larger circuit sizes. Specifically, we compare uncut circuit execution against two circuit partitioning approaches: Qiskit's automatic cut finding technique and a custom performance optimized circuit cutting method. We also measure these across GHZ, QFT, brickwork, and random quantum circuits ranging from 4 to 14 qubits, using mean absolute error of expectation values and overall output fidelity. Our results show that partitioning benefits larger, highly interconnected circuits, with our custom method reducing error by up to 55\% and improve fidelity for GHZ circuits, but degrading performance for brickwork circuits at larger scales. 
\end{abstract}

\section{Introduction}
\begin{figure*}
    \centering
    \includegraphics[width=1\linewidth]{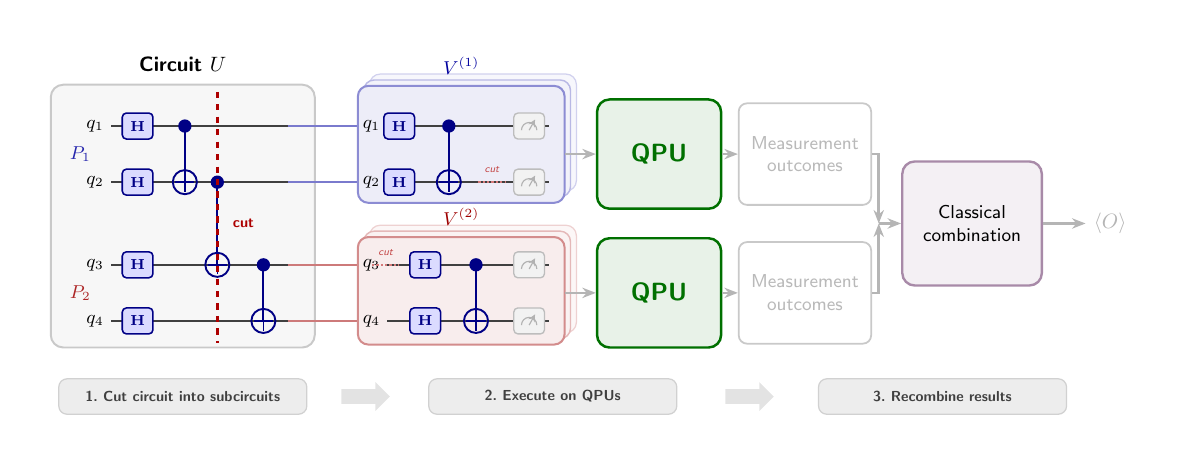}
    \caption{Quantum circuit cutting workflow. A circuit is partitioned into smaller subcircuits at selected cut locations, executed independently on available quantum processing units (QPUs), and recombined via classical post-processing to reconstruct the original circuit output. }
    \label{fig:workflow}
\end{figure*}

Quantum computing promises transformative capabilities for problems in quantum chemistry \cite{RevModPhys.92.015003}, machine learning \cite{9032461}, and combinatorial optimization \cite{Abbas2024}. Despite rapid hardware advances from IBM, Google Quantum AI, IonQ, and Quantinuum, practical quantum circuits for these applications frequently require more qubits and deeper gate sequences than current hardware can reliably execute. Even circuits with 50 to 100 qubits, though within the scope of useful algorithms, can exceed the execution capacity without fault-tolerant protection due to noise, connectivity constraints, and decoherence. 
Quantum circuit partitioning (QCP), also referred to as circuit cutting, has emerged as a practical technique to bridge this capability gap. By decomposing a large quantum circuit into smaller, independently-executable subcircuits, circuit cutting enables the distribution of quantum workloads across multiple QPU devices\cite{peng2020simulating}. The results of decomposed subcircuit executions are recombined through classical post-processing to reconstruct the output of the original circuit from its partitioned components\cite{Tang2021CutQC, peng2020simulating}. Circuit partitioning has also been demonstrated as one approach for scaling beyond single-device limitations in distributed quantum computing (DQC), where subcircuits can be executed across multiple quantum processors with classical coordination \cite{CarreraVazquez2024}. In this setting, circuit partitioning naturally aligns with modular and networked quantum architectures that would otherwise be limited by limited qubit connectivity and other hardware constraints. A schematic illustration of this workflow is shown in Fig. \ref{fig:workflow}. 

This paper presents an experimental study of QCP in the context of DQC applications. We implement and evaluate a Qiskit 2.0-compatible pipeline that integrates the FitCut community-based circuit partitioning algorithm \cite{Kan2024FitCut} with a hardware-aware qubit mapping strategy inspired by DisMap \cite{Du2024DisMap}, and the \texttt{qiskit-addon-cutting} library for subcircuit reconstruction. Our aim is to give practitioners a clear, empirically grounded understanding of where circuit partitioning is useful, how it can be implemented with modern open-source tooling, and what performance can be expected across representative circuit types and hardware configurations.

We remark that it remains unclear which classes of circuits benefit most from partitioning and under what conditions it is most effective. In particular, trade-offs between reconstruction overhead, fidelity, and execution efficiency are highly dependent on circuit structure and hardware constraints. In this work, we evaluate circuit partitioning across multiple circuit families, including GHZ, QFT, brickwork, and random circuits \cite{brandao2016local}, using mean absolute error of expectation values and output fidelity as performance metrics. We compare non-partitioned execution with both an automatic cut-finding approach and a custom performance-optimized method. Our results identify regimes in which partitioning improves fidelity and scalability, as well as cases in which it introduces performance degradation. 

The remainder of this paper is organized as follows: Section~II introduces background and related work, Section~III describes the methodology and experimental setup, Section~IV presents results and analysis, and Section~V concludes with discussion and future directions.

\subsection{Motivation for Circuit Cutting}

One of the most direct motivations for circuit cutting is enabling the execution of circuits that are either too wide in qubit count or too deep in the number of successive gate operations for a single available QPU. Cuts are introduced at selected two-qubit gates or wire segments to decompose an initial large circuit into smaller subcircuits that can be executed independently on resource-constrained, yet readily accessible hardware. The placement of these cuts is not unique and depends on the circuit topology, hardware connectivity constraints, and the chosen optimization objective, such as minimizing reconstruction overhead or preserving fidelity. The results from subcircuit executions are then recombined \cite{matsumoto2026applicabilitylimitationsquantumcircuit} through classical post-processing and, where applicable, coordinated across distributed or networked execution environments, as illustrated in Fig 1.

\subsubsection{Circuit Cutting for Parallel Circuit Execution}

A second, distinct motivation for circuit cutting is the potential to exploit parallelism in quantum circuits. Even when a circuit fits within the capacity of a single quantum processing unit (QPU), it may contain sub-computations with limited entanglement between them. In such cases, circuit cutting can be used to isolate these weakly coupled regions into subcircuits that can be executed concurrently across multiple QPUs, potentially reducing overall wall-clock execution time.

However, parallel execution does not universally guarantee speedup. Circuit cutting introduces additional sampling and reconstruction overhead, and the net performance gain depends on the balance between parallel execution benefits and these added costs. In favorable settings, particularly when inter-subcircuit dependencies are minimal, parallelization can offset this overhead and improve execution efficiency.

Parallel execution may also provide noise-related benefits. Executing a large circuit monolithically exposes all qubits to decoherence over the full circuit depth, whereas decomposing the circuit into shorter subcircuits can reduce the effective coherence time required per execution. This can help mitigate accumulated errors, although the overall impact depends on circuit structure and reconstruction complexity \cite{Tang2021CutQC}.

This use case is particularly relevant for quantum algorithms with block-structured or layered designs, such as hardware-efficient ans\"{a}tze (HWEA) used in variational quantum eigensolvers and quantum approximate optimization algorithm (QAOA) circuits whose underlying problem graphs decompose into weakly connected subgraphs. In these cases, natural partition boundaries with limited inter-subcircuit entanglement make them suitable candidates for parallel execution with manageable reconstruction overhead.
\section{Related Work}
The rapid evolution of quantum hardware has spurred significant interest in scalable, distributed quantum computing (DQC) architectures. Despite notable milestones in qubit count and gate fidelity, individual quantum processors remain constrained in qubit capacity, motivating multi-node quantum system designs that mirror the evolution of classical high-performance computing. Barral et al. review DQC architectures across varying scales \cite{barral2023distributed}, while Shapourrian et al. analyze scalable quantum data center designs based on interconnected processors \cite{shapourian2025quantum}.

On the hardware side, chip-to-chip connections have demonstrated the feasibility of distributed superconducting quantum computing. For example, IBM's Quantum Flamingo architecture connects Heron R2 processors via physical connectors measuring up to a meter in length, achieving inter-chip error rates of approximately 3.5\% \cite{Du2024DisMap}. Moreover, LaRacuente et al.\ model short-range microwave interconnects as a pathway to scale superconducting architectures further \cite{laracuente2025modeling}. On the software side, H\"{a}ner et al.\ introduced QMPI, an extension of the Message Passing Interface (MPI) for distributed quantum algorithm execution \cite{10.1145/3458817.3476172}, and Wu et al.\ proposed QuComm, a compiler framework that identifies collective communication patterns in distributed quantum programs to reduce inter-node communication by up to 54.9\% on average \cite{10.1145/3613424.3614253}.

\subsection{Quantum Circuit Cutting and Partitioning}
The theoretical foundations of wire cutting were established by Peng et al.\ \cite{peng2020simulating}, who showed that large quantum circuits can be simulated via tensor network techniques applied to results from smaller circuits. Tang et al.\ subsequently introduced CutQC \cite{Tang2021CutQC}, a comprehensive pipeline that automates cut identification using Mixed-Integer Programming (MIP) and demonstrated that hybrid quantum-classical evaluation can produce lower-noise outputs than purely quantum execution on analogously sized circuits. However, CutQC's cut searcher faces factorial search complexity in the number of circuit edges, resulting in runtimes of 100 to 1000 seconds for large circuits.

To address this scalability bottleneck, Kan et al.\ proposed FitCut \cite{Kan2024FitCut}, a community-based bottom-up approach that transforms the circuit DAG into a gate-only weighted graph and applies constrained modularity-based community detection \cite{Blondel2008} to identify partition boundaries. FitCut incorporates a two-tiered objective that jointly minimizes the number of cuts and maximizes qubit utilization across heterogeneous worker configurations, reducing cutting time by 3 to 2000$\times$ over the Qiskit Circuit Knitting Toolbox while improving per-worker resource utilization by up to 388\%. Brandhofer et al.\ \cite{brandhofer2023optimal} approached optimal partitioning via a Satisfiability Modulo Theories (SMT) model considering both gate cuts and wire cuts, and Andr\'{e}s-Mart\'{i}nez and Heunen \cite{Andr_s_Mart_nez_2019} framed circuit distribution as a hypergraph partitioning problem, where qubits map to vertices and QPUs map to blocks such that minimizing cut hyperedges corresponds to minimizing entanglement resource usage between processors.

\subsection{Hardware-Aware Compilation and Qubit Mapping}
Realizing circuit cutting in practice requires hardware-aware compilation to ensure subcircuits are executable on target QPUs. Du et al.\ proposed DisMap \cite{Du2024DisMap}, a self-adaptive framework for chip-to-chip distributed quantum systems that analyzes qubit noise levels and error rates to construct a Virtual System Topology (VST). DisMap uses this topology to guide both circuit partitioning (via FitCut) and distributed subcircuit qubit mapping, minimizing SWAP overhead while maximizing fidelity. Evaluated on IBM-Q hardware emulators, DisMap achieves up to 20.8\% improvement in fidelity and reduces SWAP overhead by up to 80.2\% compared to Sabre-based baselines \cite{Li2019Sabre}.

Error mitigation is an essential companion to circuit cutting in the NISQ era. The Mitiq library \cite{mitiq} provides Richardson extrapolation and probabilistic error cancellation applicable to subcircuit results prior to reconstruction. Ioannou et al.\ \cite{Ioannou2024CanMR} further demonstrate that machine learning-based reliability models can predict circuit fidelity across hardware backends, offering a pathway toward adaptive cut placement that accounts for time-varying hardware noise profiles.

\subsection{Applications of Circuit Cutting}
In addition to improving scalability, circuit cutting enables execution models aligned with distributed and networked quantum systems by breaking a large circuit into smaller components that can be run independently on separate quantum processing units and later combined through classical coordination \cite{Tang2021CutQC}. This approach is particularly well suited for modular architectures, where limited qubit connectivity prevents direct realization of large, globally entangled circuits. From a cybersecurity point of view, distributing execution across subcircuits can also limit exposure of the entire circuit structure in shared or cloud-based environments \cite{szefer2024leveraging}, although security considerations are not the main focus of this work.

\section{Background}
A central limitation of current quantum hardware is \emph{decoherence}. Interactions with the environment degrade quantum states over timescales $T_1$ (relaxation) and $T_2$ (dephasing). Gate operations also introduce errors, with two-qubit gates typically exhibiting higher error rates than single-qubit gates. 

\subsection{IBM Qiskit}
This project uses Qiskit~~\cite{qiskit2024} version 2.0, which introduces a redesigned transpiler API centered on the \texttt{StagedPassManager}, replacing the legacy \texttt{transpile()} call with a configurable pipeline of optimization passes. The \texttt{qiskit-addon-cutting} library extends Qiskit with quasi-probability decomposition (QPD) based circuit cutting and subcircuit reconstruction, and serves as the reconstruction backbone of our pipeline.

\subsection{Quantum Circuit Cutting}
\subsubsection{Wire Cuts and Gate Cuts}
There are two primary mechanisms by which a quantum circuit can be cut. A \textbf{wire cut} interrupts the flow of quantum information along a qubit wire between two gates. The cut is handled by replacing the severed wire with a set of state preparations and measurements, effectively converting the cross-cut quantum channel into a classically communicated probability distribution. A \textbf{gate cut} decomposes a multi-qubit gate, such as a CNOT, into a sum of local single-qubit operations on either side of the cut, eliminating the need for the qubits to interact during execution \cite{brandhofer2023optimal}. This work employs wire cuts implemented via the \texttt{qiskit-addon-cutting} library.

\subsubsection{Quasi-Probability Decomposition and Reconstruction}
The theoretical basis for both cut types is \emph{quasi-probability decomposition} (QPD). A non-local quantum channel $\mathcal{E}$ across a cut is expressed as a weighted sum of local, implementable channels $\{\mathcal{F}_i\}$:
\begin{equation}
    \mathcal{E} = \sum_i c_i \, \mathcal{F}_i, \quad c_i \in \mathbb{R}, 
\end{equation}
where the coefficients $c_i$ may be negative, hence the term \emph{quasi}-probability. The full circuit's expectation value for an observable $O$ is then recovered by running each local variant $\mathcal{F}_i$ independently and combining the results classically:
\begin{equation}
    \langle O \rangle = \sum_i c_i \langle O \rangle_{\mathcal{F}_i}.
\end{equation}
The reconstructed probability distribution of the full circuit can also be expressed as a classical sum over intermediate measurement outcomes at the cut points \cite{peng2020simulating}:
\begin{equation}
    P_\mathrm{full}(x) = \sum_z P_1(x \,|\, z)\, P_2(z),
\end{equation}
where $z$ represents intermediate states at the cut and $P_1$, $P_2$ are the conditional probability distributions from the respective subcircuits.

\subsubsection{Sampling Overhead}
QPD introduces a \emph{sampling overhead}: correctly estimating the expectation values from the decomposition requires exponentially more circuit shots than executing the uncut circuit. For $n$ cuts, the sampling overhead scales as $\mathcal{O}(4^n)$ \cite{Tang2021CutQC}. This scaling is the primary practical limitation of circuit cutting and provides the core motivation for minimizing the number of cuts, which is the primary objective of FitCut.

\subsection{Distributed Quantum Computing}
Distributed quantum computing (DQC) refers to the use of multiple interconnected quantum processors to collaboratively execute circuits that exceed the capacity of any single device. Two principal connection paradigms are commonly considered. \emph{Quantum network} approaches connect quantum processing units (QPUs) over distance through the distribution of entanglement, enabling applications such as quantum key distribution. In contrast, \emph{chip-to-chip} connections integrate multiple quantum chips within a single system or across short distances, allowing direct qubit interactions via shared physical links or microwave connectors \cite{barral2023distributed}.

Central to chip-to-chip DQC is the Einstein-Podolsky-Rosen (EPR) pair: a maximally entangled two-qubit state $\frac{1}{\sqrt{2}}(\lvert 00\rangle + \lvert 11\rangle)$ shared between qubits on separate chips. EPR pairs serve as the resource for non-local gate operations, teleportation, and entanglement swapping across processors. Establishing an EPR pair introduces additional noise due to decoherence during entanglement generation, making the choice of which physical qubit pairs to entangle---and across which cut boundaries---a critical optimization decision \cite{Du2024DisMap}.

\section{Methodology}
\label{sec:methodology}

We study circuit cutting within the standard cut-and-reconstruct framework, in which a quantum circuit is decomposed into smaller subcircuits at selected two-qubit interaction points and the target observables are recovered through classical post-processing of subexperiment outcomes. This framework is based on quasiprobability decomposition (QPD), where nonlocal operations are expressed as weighted combinations of locally executable ones \cite{Temme2017, Piveteau2021}. Circuit cutting therefore enables observable estimation under hardware qubit and connectivity constraints, but at the cost of sampling overhead that grows rapidly with the number of cuts. Consequently, practical performance depends critically on how cut locations and observables are selected.

Within this framework, we compare two cut-selection methods, Qiskit's automatic cut-finding method and a custom budget-aware heuristic, \texttt{fitv3}, together with a direct no-cut baseline. Qiskit's method follows the circuit-knitting paradigm, where cut locations are identified automatically subject to a maximum subcircuit-width constraint and expectation values are reconstructed through quasiprobability sampling \cite{Piveteau2021, Schmitt2025}. The \texttt{fitv3} method, by contrast, is a custom FitCut-inspired \cite{Kan2024FitCut, Du2024DisMap} heuristic that searches over candidate gate-cut sets and scores them using structural and budget-aware criteria, including partition size, subexperiment burden, and estimated sampling overhead. Because all methods operate under the same simulator, observable set, and reconstruction framework, the comparison isolates how cut-selection strategy affects observable accuracy, feasibility, and practical reliability.

To evaluate performance, we focus on observable-level accuracy rather than full-state reconstruction. Let $C$ denote the input circuit and let $\mathcal{O} = \{O_1,\dots,O_k\}$ be the observable set used for evaluation. For each circuit instance, we first remove terminal measurements to obtain a nominal circuit $C_{\mathrm{nom}}$, and then compute ideal expectation values from the noiseless statevector of $C_{\mathrm{nom}}$:
\begin{equation}
E_{\mathrm{ideal}}(O_i) = \langle \psi_C \mid O_i \mid \psi_C \rangle,
\end{equation}
where $\lvert \psi_C \rangle$ is the statevector obtained from $C_{\mathrm{nom}}$.

Each strategy then produces an expectation value $\hat{E}(O_i)$ for every observable $O_i \in \mathcal{O}$. For the direct no-cut baseline, these estimates come from noisy execution of the original circuit. For the cut-based methods, they are obtained by generating subexperiments, executing them under the same noisy simulator, and recombining the results through the \texttt{qiskit-addon-cutting} reconstruction pipeline. At the observable level, QPD reconstruction can be written as
\begin{equation}
\hat{E}(O) = \sum_j c_j \, \hat{E}_j(O),
\end{equation}
where $c_j$ are the quasiprobability coefficients and $\hat{E}_j(O)$ is the sampled estimate from the $j$th generated local experiment.

For an observable $O$, the reconstruction error is defined as
\begin{equation}
\Delta_O = \left| \hat{E}(O) - E_{\mathrm{ideal}}(O) \right|.
\end{equation}
Aggregating across the observable set, we use the mean absolute error (MAE) as the primary accuracy metric:
\begin{equation}
\mathrm{MAE} = \frac{1}{k} \sum_{i=1}^{k}
\left| \hat{E}(O_i) - E_{\mathrm{ideal}}(O_i) \right|.
\end{equation}
This metric is matched to the circuit-cutting setting, since the goal of reconstruction in the present study is accurate recovery of selected observable expectation values rather than the full-states.

To compare each strategy directly against the no-cut baseline, we use the following relation:
\begin{equation}
\Delta \mathrm{MAE} =
\mathrm{MAE}_{\texttt{method}} - \mathrm{MAE}_{\texttt{no\_cut}}.
\end{equation}
Negative values therefore indicate that a strategy improves upon direct execution, while positive values indicate worse reconstruction accuracy than the no-cut baseline. For repeated runs at fixed circuit width, we also report the win rate of \texttt{fitv3} relative to the no-cut baseline:
\begin{equation}
\mathrm{WinRate}(n) =
\frac{1}{R}\sum_{r=1}^{R}
\mathbf{1}\!\left[
\mathrm{MAE}_{\texttt{fitv3}}^{(r,n)}
<
\mathrm{MAE}_{\texttt{no\_cut}}^{(r,n)}
\right],
\end{equation}
where $R$ is the number of repeated runs at qubit count $n$ and $\mathbf{1}[\cdot]$ is the indicator function.

\subsection{Setup}

All strategies begin from the same nominal circuit $C_{\mathrm{nom}}$ and the same ideal observable values. The no-cut baseline produces direct noisy estimates of the target observables. The cut-based methods instead generate subcircuits, execute the corresponding subexperiments on the same noisy simulator, and reconstruct the final observable estimates using \texttt{qiskit-addon-cutting}. Because the execution and reconstruction pipeline is otherwise shared, differences in performance can be attributed to the cut-selection strategy itself.

We evaluate three circuit families: GHZ, QFT, and a random circuit family with alternating nearest-neighbor entangling layers and randomized single-qubit operations. For the first sweep, we use the \texttt{z\_magnetization} observable family, which defines a single observable equal to the average of local $Z$ operators across all qubits:
\begin{equation}
O_{\mathrm{mag}} = \frac{1}{n}\sum_{i=1}^{n} Z_i.
\end{equation}
In additional GHZ-specific experiments, we also evaluate the \texttt{ghz\_stabilizers} observable set.

All methods are evaluated on the same circuit instances, observable sets, simulator profile, and sampling parameters. The no-cut baseline is also evaluated under the same noisy simulator used by the cut-based methods. When the observable set is $Z$-only, the baseline uses a counts-based expectation path so that all methods are compared through finite-shot sampled estimates rather than mixing exact and sampled expectation evaluation.

Not every candidate cut configuration is accepted. A run may be skipped if the proposed decomposition violates the maximum subcircuit-width constraint or if its estimated sampling overhead exceeds the configured budget threshold. In addition to MAE, we therefore record skip indicators and associated reasons for each run, along with method-specific metadata such as the number of cuts, estimated sampling overhead, generated subexperiment count, and total shots.

\subsection{Qiskit Auto-Cut Algorithm}

\begin{algorithm}[t]
\small
\caption{Qiskit auto cutting}
\label{alg:qiskit-auto}
\begin{algorithmic}[1]
\Require Circuit $C$, observables $\mathcal{O}$, width limit $q_{\max}$, overhead cap $\Gamma_{\max}$
\Ensure Reconstructed expectations $\{\hat{E}(O_i)\}$ or skip
\State $C_{\mathrm{nom}} \gets$ remove measurements and decompose to 1Q/2Q gates
\State Compute ideal expectations $\{E_{\mathrm{ideal}}(O_i)\}$ from $C_{\mathrm{nom}}$
\State $\mathcal{F} \gets \emptyset$
\For{each preset search configuration $\theta$}
    \State $(C_\theta,D_\theta) \gets \texttt{find\_cuts}(C_{\mathrm{nom}}, q_{\max}, \theta)$
    \If{$(C_\theta,D_\theta)$ is feasible and $\Gamma_\theta \le \Gamma_{\max}$}
        \State $\mathcal{F} \gets \mathcal{F} \cup \{(C_\theta,D_\theta)\}$
    \EndIf
\EndFor
\If{$\mathcal{F}=\emptyset$}
    \State \Return skip
\EndIf
\State Select best feasible candidate $(C^\star,D^\star)$
\State Generate subexperiments and QPD coefficients from $C^\star$
\State Execute subexperiments and reconstruct $\hat{E}(O_i)$ for all $O_i \in \mathcal{O}$
\State \Return $\{\hat{E}(O_i)\}, \{E_{\mathrm{ideal}}(O_i)\}$
\end{algorithmic}
\end{algorithm}

The automatic baseline uses Qiskit's addon cut finder, but in our implementation it is not treated as a single black-box call. Instead, several preset search configurations are attempted, each candidate is screened against the subcircuit-width constraint and the global sampling-overhead cap, and candidate diagnostics are retained even when no feasible cut set is found. This makes the automatic method a fair practical baseline under the same reconstruction budget used by the custom heuristic.

\subsection{Fitv3 Custom Method Algorithm}

\begin{algorithm}[t]
\small
\caption{fitv3 budget-aware cut selection}
\label{alg:fitv3}
\begin{algorithmic}[1]
\Require Circuit $C$, observables $\mathcal{O}$, cut budget $c_{\max}$, width limit $q_{\max}$, overhead cap $\Gamma_{\max}$
\Ensure Reconstructed expectations $\{\hat{E}(O_i)\}$
\State $C_{\mathrm{nom}} \gets$ remove measurements from $C$
\State Compute ideal expectations $\{E_{\mathrm{ideal}}(O_i)\}$ from $C_{\mathrm{nom}}$
\State Build candidate cut pool $P$ from two-qubit gate locations
\State $\mathcal{K} \gets \{K \subseteq P : |K| \le c_{\max}\}$
\State $K^\star \gets \emptyset,\; S^\star \gets -\infty$
\For{each $K \in \mathcal{K}$}
    \State Estimate partition sizes, subexperiment count, and overhead $\Gamma(K)$
    \If{$\Gamma(K) \le \Gamma_{\max}$}
        \State Compute score $S(K)$
        \If{$S(K) > S^\star$}
            \State $K^\star \gets K,\; S^\star \gets S(K)$
        \EndIf
    \EndIf
\EndFor
\State Apply $K^\star$ to obtain cut circuit $C^\star$ (or keep $C_{\mathrm{nom}}$ if $K^\star=\emptyset$)
\State Generate subexperiments and QPD coefficients from $C^\star$
\State Execute subexperiments and reconstruct $\hat{E}(O_i)$ for all $O_i \in \mathcal{O}$
\State \Return $\{\hat{E}(O_i)\}, \{E_{\mathrm{ideal}}(O_i)\}$
\end{algorithmic}
\end{algorithm}

The custom \texttt{fitv3} method is a budget-aware gate-cut heuristic rather than a direct implementation of the original FitCut community-detection algorithm \cite{Kan2024FitCut}. It first extracts the two-qubit interaction structure of the circuit, proposes candidate cut locations using simple partition templates, and then evaluates candidate cut sets up to a fixed cut budget. Candidates are scored using execution-aware criteria that balance observable support, cut placement, partition size, generated subexperiment count, and estimated overhead. A candidate with $n_{\mathrm{cuts}}$ selected gate cuts is assigned an estimated sampling overhead
\begin{equation}
\Gamma(K) = 9^{\,n_{\mathrm{cuts}}},
\end{equation}
and candidates whose estimated overhead exceeds the configured cap are rejected. The selected cut set is therefore intended not merely to satisfy the width constraint, but to remain useful under finite sampling budgets.

\subsection{Environment Configuration}

\subsubsection{Software Environment}
All experiments were run in Python 3.11 using Qiskit, Qiskit Aer, and \texttt{qiskit-addon-cutting}. The evaluation was executed in a Google Colab environment.

\subsubsection{Simulator Configuration}
We evaluate all methods under a common noisy simulation setting derived from IBM's \texttt{ibm\_brisbane} hardware profile. Specifically, we use the corresponding fake backend to obtain the device coupling map, basis-gate information, and backend noise properties, and then construct a Qiskit Aer noise model from that profile. This allows all methods to be compared under the same hardware-inspired simulator rather than under idealized noiseless execution.

\subsubsection{Backend Profile}
The simulator is based on the \texttt{ibm\_brisbane} backend profile available at evaluation time. Because backend calibration data can drift, we treat this profile as a hardware-inspired reference configuration rather than a claim about a single immutable device snapshot.

\section{Evaluation}

We evaluate three execution strategies for observable estimation: direct uncut execution, Qiskit's automatic cut-finding method, and a custom budget-aware heuristic method. All experiments are run under a common noisy simulation setting derived from IBM's Brisbane hardware profile using Qiskit Aer and the corresponding fake backend configuration. The goal of the study is to identify when cutting remains competitive with direct execution and when it becomes unreliable under a realistic noise profile.

The benchmark families are GHZ, QFT, and random circuits.. Circuit widths range from 4 to 16 qubits. For the main sweep, we use the \texttt{z\_magnetization} observable family for all circuit classes, and we repeat each configuration 10 times with different seeds. Our fixed cutting constraints are a maximum subcircuit width of 4 qubits, 200 shots per subexperiment, 100 reconstruction samples, and a maximum allowed sampling overhead of $10^8$. For the custom method, we fix the cut budget to at most two cuts and use the final \texttt{fitv3} configuration throughout.

For every run, we record the ideal expectation values from the noiseless statevector of the unmeasured circuit and compare them against the reconstructed expectation values returned by each strategy. Our primary accuracy metric is the mean absolute error (MAE),
\begin{equation}
\mathrm{MAE} = \frac{1}{k}\sum_{i=1}^{k} \left| \hat{E}(O_i) - E_{\mathrm{ideal}}(O_i) \right|,
\end{equation}
where $\hat{E}(O_i)$ is the reconstructed expectation value for observable $O_i$. We also record whether a strategy was skipped or rejected due to infeasibility or overhead limits, along with diagnostic quantities such as the number of cuts, estimated sampling overhead, number of generated subexperiments, and total shots.

\subsection{Circuit Family Results}

\begin{figure*}[t]
\centering

\begin{subfigure}{0.48\textwidth}
    \includegraphics[width=\linewidth]{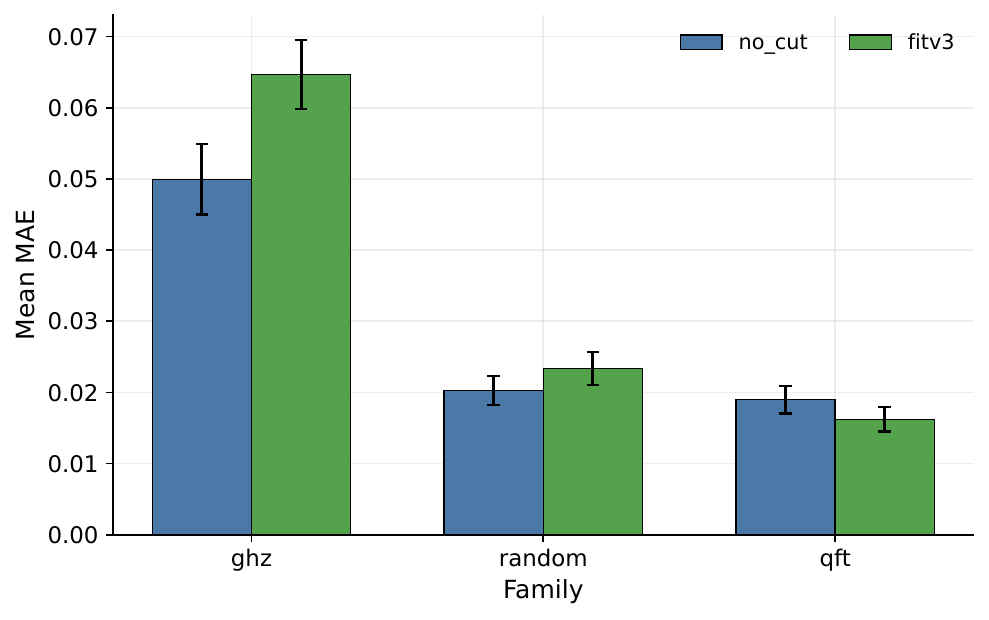}
    \caption{Family-level mean MAE for the direct no-cut baseline and the custom fitv3 cutting heuristic. Lower values are better.}
    \label{fig:family-stable}
\end{subfigure}
\hfill
\begin{subfigure}{0.48\textwidth}
    \includegraphics[width=\linewidth]{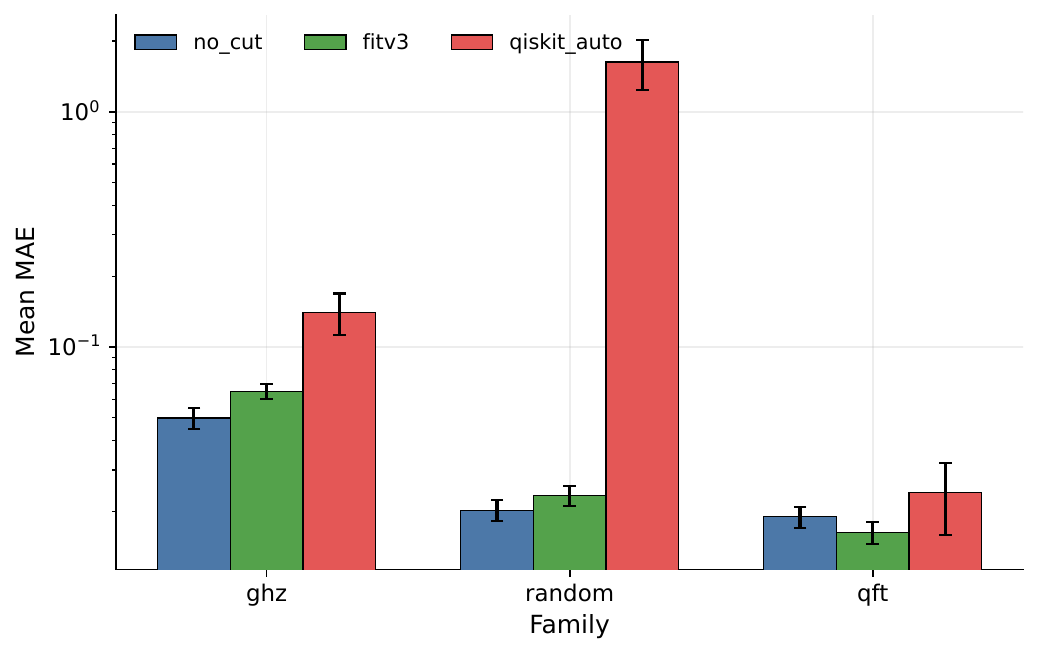}
    \caption{Family-level mean MAE for all three strategies on a log scale. This view makes the full spread visible, including the much larger errors produced by Qiskit auto on the random family.}
    \label{fig:family-log}
\end{subfigure}

\begin{subfigure}{0.48\textwidth}
    \includegraphics[width=\linewidth]{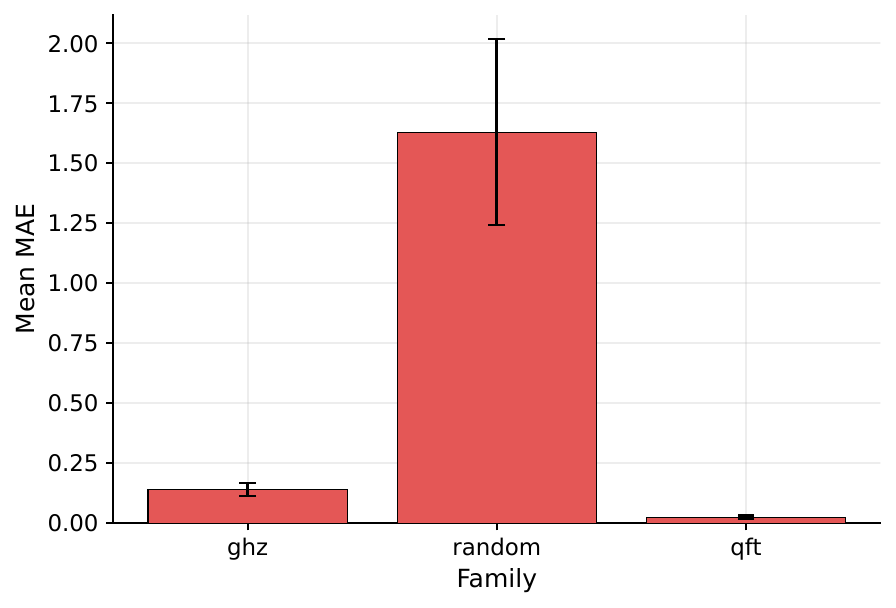}
    \caption{Family-level mean MAE for Qiskit's automatic cut-finding method.}
    \label{fig:qiskit-auto-family}
\end{subfigure}
\hfill
\begin{subfigure}{0.48\textwidth}
    \includegraphics[width=\linewidth]{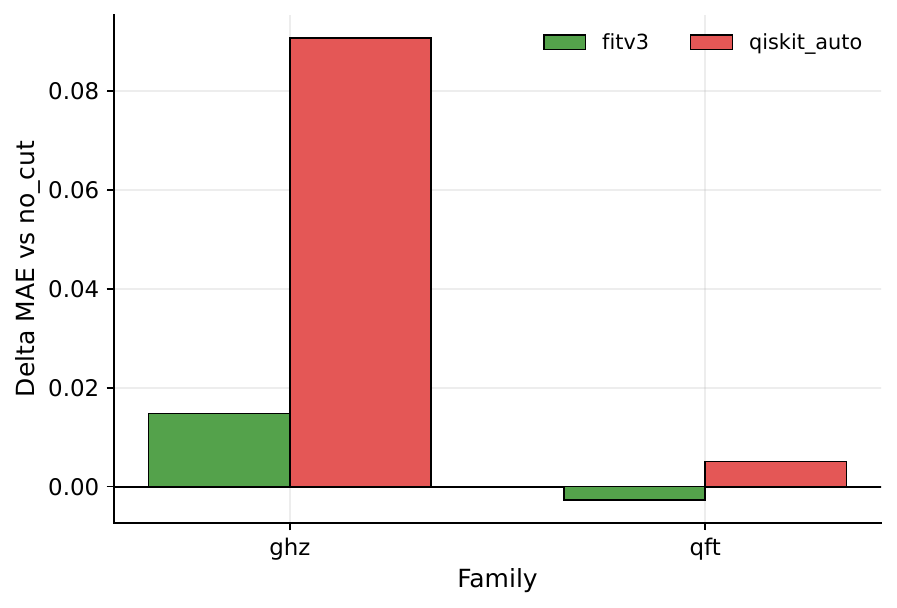}
    \caption{Family-level $\Delta \mathrm{MAE}$ relative to the no-cut baseline for GHZ and QFT.}
    \label{fig:delta-family}
\end{subfigure}
\label{fig:combined-2x2}
\end{figure*}

Figure~\ref{fig:family-stable} compares the family-level mean MAE of the direct \texttt{no\_cut} baseline against \texttt{fitv3}. Across GHZ, random, and QFT circuits, \texttt{fitv3} remains close to the no-cut baseline, indicating that the custom heuristic usually avoids the catastrophic degradation associated with poor cut placement. This is an important result: a practically useful cutting strategy does not need to dominate direct execution everywhere, but it must remain stable enough that cutting does not introduce arbitrarily large reconstruction error.

Figure~\ref{fig:qiskit-auto-family} isolates the family-level behavior of \texttt{qiskit\_auto}. In contrast to \texttt{fitv3}, Qiskit's automatic method performs especially poorly on random circuits, where feasible cut sets can still yield very large MAE. This indicates that cut feasibility alone is not a sufficient success criterion. A returned partition may satisfy the subcircuit-width constraint and still be scientifically poor due to large reconstruction cost.

\begin{figure}[t]
    \centering
    \includegraphics[width=\linewidth]{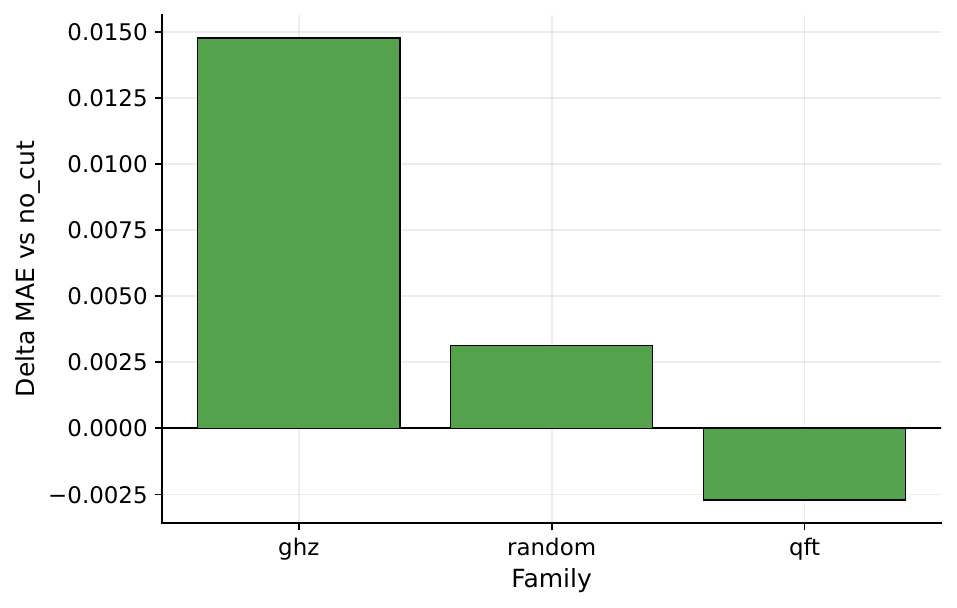}
    \caption{Family-level $\Delta \mathrm{MAE}$ for fitv3 relative to the no-cut baseline across all three circuit families. Positive values indicate worse performance than direct execution, while negative values indicate improvement.}
    \label{fig:delta-fitv3-only}
\end{figure}

To compare the cut methods directly against the no-cut baseline, Figure~\ref{fig:delta-family} reports family-level $\Delta \mathrm{MAE}$ relative to \texttt{no\_cut} for GHZ and QFT. Positive values indicate worse performance than direct execution, while negative values indicate improvement. The custom method remains much closer to zero than \texttt{qiskit\_auto}, and on QFT it is slightly favorable on average. This supports the interpretation that the custom method is the more stable cutting baseline, while \texttt{qiskit\_auto} is primarily useful as a negative comparison baseline.

\subsection{Size-Dependent Comparison to No-Cut}
\begin{figure}[t]
    \centering
    \includegraphics[width=\linewidth]{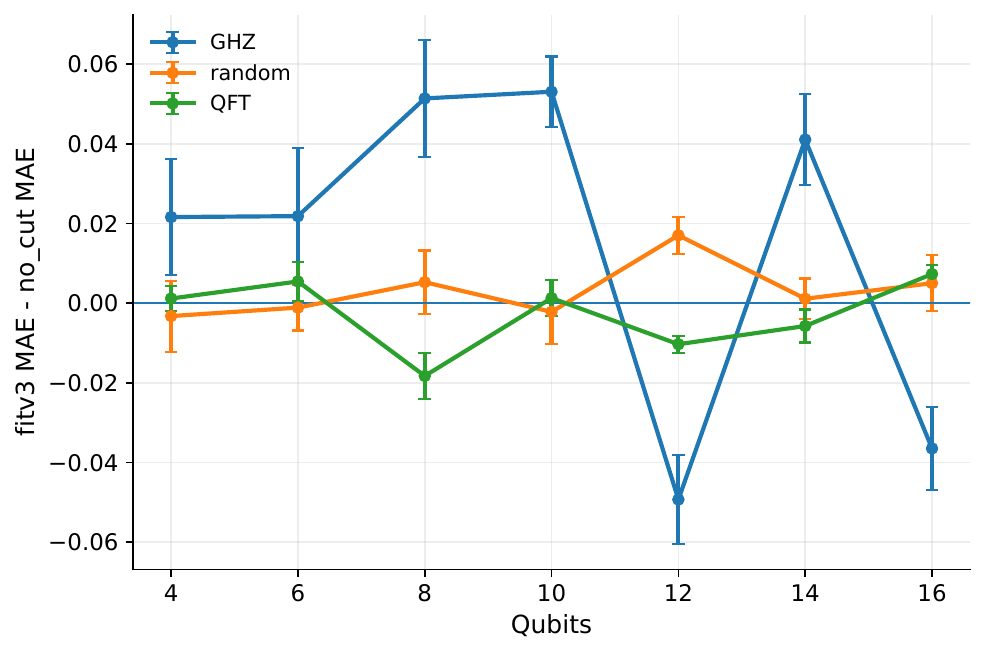}
    \caption{Difference in MAE between fitv3 and the no-cut baseline as a function of qubit count for each circuit family.}
    \label{fig:delta-by-qubits}
\end{figure}

Aggregate family averages hide an important part of the story.Cutting is not uniformly good or bad, but rather depends on both circuit structure and size. To expose this behavior, Figure~\ref{fig:delta-by-qubits} plots the difference
\begin{equation}
\Delta \mathrm{MAE}(n) = \mathrm{MAE}_{\texttt{fitv3}}(n) - \mathrm{MAE}_{\texttt{no\_cut}}(n)
\end{equation}
as a function of qubit count for each family. Negative values indicate that cutting improves upon direct execution.

The results show that \texttt{fitv3} is not a universal replacement for no-cut execution. For GHZ circuits, it underperforms the no-cut baseline at several smaller and medium sizes, but becomes favorable at selected larger instances. For QFT circuits, it remains competitive across most of the sweep and outperforms no-cut at multiple sizes. For random circuits, the difference stays close to zero overall, indicating near-parity rather than a strong consistent gain in either direction. Thus, the value of cutting is most visible on structured families and in specific size range rather than as a blanket improvement.

\begin{figure}[t]
    \centering
    \includegraphics[width=\linewidth]{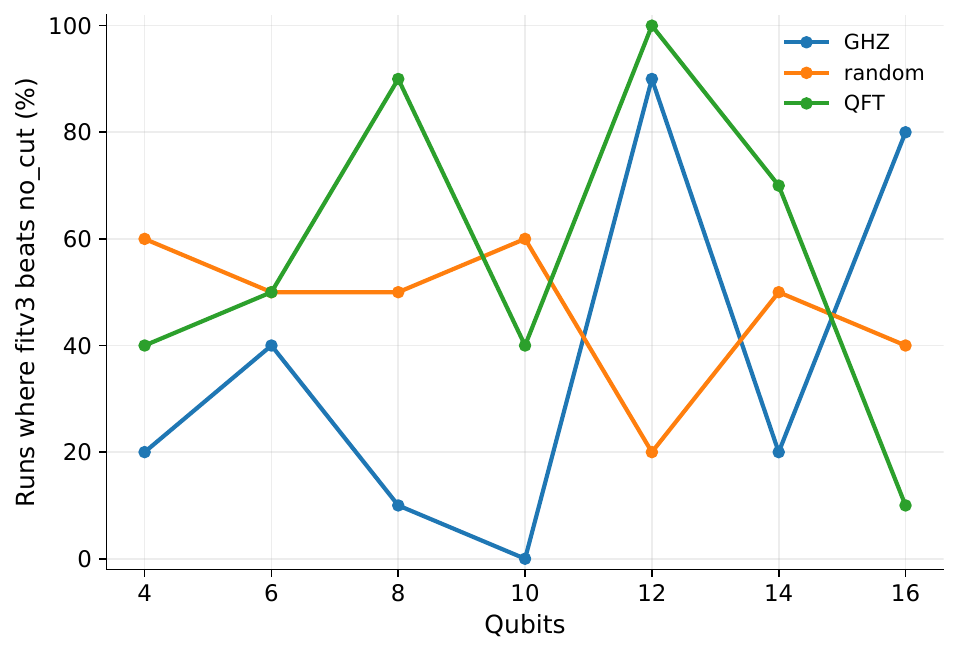}
    \caption{Fraction of repeated runs for which fitv3 achieves lower MAE than the no-cut baseline as a function of qubit count for each circuit family.}
    \label{fig:winrate-by-qubits}
\end{figure}

Figure~\ref{fig:winrate-by-qubits} complements the average-error view with a per-size win-rate. For each family and qubit count, the figure reports the fraction of matched repeated runs in which \texttt{fitv3} achieved lower MAE than \texttt{no\_cut}. This plot is useful because a small average $\Delta \mathrm{MAE}$ can conceal a mixed result in which cutting sometimes helps and sometimes hurts. The win-rate results show that \texttt{fitv3} wins frequently for selected larger GHZ and QFT cases, while remaining mixed on random circuits. In other words, the custom method is most promising on structured circuits where low-cost cuts can capture meaningful separation without triggering excessive reconstruction overhead.

\subsection{Failure Modes of Automatic Cut Finding}

The weakest strategy in the study is \texttt{qiskit\_auto}, but its failure is not one-dimensional. Our results reveal two distinct failure modes. First, the automatic routine can reject a candidate because the estimated sampling overhead exceeds the configured cap. Second, even when it returns a feasible structure, the resulting reconstruction can still be poor because the cut configuration generates too many subexperiments or too much effective variance. This behavior is explicitly reflected in the results, which store skip flags, reasons, candidate diagnostics, and estimated cut overhead for each run.

Empirically, this instability is most visible on the random family. The family-level MAE for \texttt{qiskit\_auto} is dramatically worse there than for either \texttt{no\_cut} or \texttt{fitv3}, indicating that the method often finds formally admissible but scientifically weak cut sets. This is exactly the type of behavior that motivates budget-aware heuristics: a good cutting method must optimize not just for structural feasibility, but for the quality of the reconstructed observable estimates.

\subsection{Summary of Findings}

Taken together, the results support three main conclusions. First, direct \texttt{no\_cut} execution remains the strongest baseline on average whenever it is feasible. Second, the custom \texttt{fitv3} heuristic is substantially more stable than Qiskit's automatic cut finder and can be competitive with, or occasionally better than, direct execution on selected larger structured circuits, especially QFT and some GHZ cases. Third, automatic cut finding can fail in two ways: by becoming infeasible under the sampling-overhead budget, or by producing feasible but high-error decompositions. The main takeaway is that circuit cutting should be evaluated through observable accuracy under explicit reconstruction budgets, not just through the existence of a valid partition.

\section{Conclusion}
    In this work, we translated the FitCut circuit partitioning algorithm \cite{Kan2024FitCut} from Qiskit 1.x to Qiskit 2.0, adapting the DAG-to-weighted-graph transformation and community detection pipeline to the redesigned Qiskit 2.0 transpiler API. We also implemented a Virtual System Topology (VST) construction procedure inspired by DisMap \cite{Du2024DisMap} that selects low-noise inter-chip qubit pairs from IBM backend calibration data and uses the resulting unified coupling graph to guide noise-aware subcircuit routing. 
    We integrated FitCut-based partitioning, VST-guided transpilation via \texttt{generate\_preset\_pass\_manager}, and QPD-based expectation value reconstruction via \texttt{qiskit-addon-cutting} into a unified, reproducible pipeline targeting the \texttt{ibm\_brisbane} hardware profile. 
    Finally, we evaluated the pipeline on representative benchmark circuits---including BV, ADDER, HWEA, and QFT variants---reporting fidelity and SWAP overhead under both noise-model simulation and execution on real IBM quantum hardware. This work demonstrates a practical approach for noise-aware circuit partitioning and execution on current quantum hardware and provides a reproducible framework for future algorithm benchmarking. 

\subsection{Future Directions}
    Several directions present themselves as natural extensions of this work. First, the current pipeline employs only wire cuts; integrating gate cuts via quasi-probability decomposition of multi-qubit unitaries \cite{brandhofer2023optimal} could reduce the required number of cuts for circuits with dense two-qubit gate interactions, thereby lowering sampling overhead. Second, the evaluation is presently limited to IBM superconducting hardware; extending the pipeline to trapped-ion (IonQ, Quantinuum) and neutral-atom backends would require adapting the VST construction and transpilation stages to different native gate sets and connectivity models, and would broaden the conclusions about where circuit cutting is practically beneficial.

    Third, the VST construction currently selects EPR link qubits from a static snapshot of backend calibration data. In practice, qubit noise profiles drift substantially over hours and days \cite{Ioannou2024CanMR}; an adaptive variant that re-queries calibration data at job submission time and re-optimizes link qubit selection accordingly could yield more consistent fidelity gains. Finally, as IBM and other vendors push toward larger qubit counts---with devices exceeding 1000 qubits already available on cloud platforms---the scalability of the FitCut community detection step and the QPD reconstruction overhead will become increasingly important to characterize, motivating future evaluation at significantly larger circuit scales.

\section*{Acknowledgment}
    This work was performed as part of WISER Solutions Launchpad program, with the first three authors listed as the WISER Research fellows.

    This manuscript has been coauthored by the Naval Nuclear Laboratory, operated by Fluor Marine Propulsion, LLC under contract No. 89233018CNR000004 with the U.S. Department of Energy. The United States Government retains and the publisher, by accepting the article for publication, acknowledges that the United States Government retains a non-exclusive, paid-up, irrevocable, world-wide license to publish, distribute, translate, duplicate, exhibit, and perform the published form of this manuscript, or allow others to do so, for United States Government purposes.

\bibliographystyle{acm}
\bibliography{references}

\end{document}